\documentclass{emulateapj}
\usepackage{amsmath}
\usepackage{graphicx}
\usepackage{natbib}
\usepackage[scaled]{helvet}
\usepackage{epsfig}
\usepackage{url}
\bibpunct{(}{)}{;}{a}{}{,}
\newcommand{\kms}{\,km\,s$^{-1}$}
\usepackage{textcomp}
\usepackage{mathpazo}
\usepackage{xspace}
\usepackage{apjfonts}

\usepackage{color}
\usepackage[normalem]{ulem}

\newcommand{\bjdtdb}{\ensuremath{\rm {BJD_{TDB}}}}
\newcommand{\feh}{\ensuremath{\left[{\rm Fe}/{\rm H}\right]}}
\newcommand{\mh}{\ensuremath{\left[{\rm m}/{\rm H}\right]}}
\newcommand{\teff}{\ensuremath{T_{\rm eff}}\xspace}
\newcommand{\logg}{\ensuremath{\log g}}

\newcommand{\msun}{\ensuremath{\,M_\Sun}}
\newcommand{\rsun}{\ensuremath{\,R_\Sun}}

\newcommand{\rearth}{\ensuremath{\,R_{\rm \Earth}}\xspace}
\newcommand{\mearth}{\ensuremath{\,M_{\rm \Earth}}\xspace}
\newcommand{\re}{\ensuremath{\,R_{\rm \Earth}}\xspace}
\newcommand{\me}{\ensuremath{\,M_{\rm \Earth}}\xspace}
\newcommand{\fave}{\langle F \rangle}
\newcommand{\fluxcgs}{10$^9$ erg s$^{-1}$ cm$^{-2}$}

\newcommand{\Kepler}{{\it Kepler}}
\newcommand{\ms}{\,m\,s$^{-1}$}
\newcommand{\minus}{\scalebox{0.75}[1.0]{$-$}}
\newcommand{\thisstar}{GJ~9827\xspace}
\newcommand{\mstar}{\ensuremath{M_{*}}}
\newcommand{\rstar}{\ensuremath{R_{*}}}
\newcommand{\ar}{\ensuremath{a/R_*}}
\newcommand{\radiusb}{$1.62\pm0.11$\xspace\re\xspace}
\newcommand{\radiusc}{$1.269^{+0.087}_{-0.089}$\xspace\re\xspace}
\newcommand{\radiusd}{$2.07\pm0.14$\xspace\re\xspace}

\begin{document}

\title{A System of Three Super Earths Transiting the Late K-Dwarf GJ 9827 at Thirty Parsecs}
\author{Joseph E. Rodriguez$^1$, Andrew Vanderburg$^{2,1,\star}$, Jason D. Eastman$^{1}$, Andrew W. Mann$^{2,3,\dagger}$, Ian J. M. Crossfield$^4$, \\ David R. Ciardi$^5$, David W. Latham$^{1}$, Samuel N. Quinn$^{1}$}

\affil{$^{1}$Harvard-Smithsonian Center for Astrophysics, 60 Garden St, Cambridge, MA 02138, USA}
\affil{$^{2}$Department of Astronomy, The University of Texas at Austin, Austin, TX 78712, USA}
\affil{$^{3}$Department of Astronomy, Columbia University, 550 West 120th Street, New York, NY 10027, USA}
\affil{$^{4}$Department of Physics, Massachusetts Institute of Technology, Cambridge, MA, USA}
\affil{$^{5}$NASA Exoplanet Science Institute, California Institute of Technology, Pasadena, CA, USA}
\affil{$^{\star}$NASA Sagan Fellow}
\affil{$^{\dagger}$NASA Hubble Fellow}

\shorttitle{Three Planets --- Thirty Parsecs}
\shortauthors{Rodriguez et al.}

\begin{abstract}
We report the discovery of three small transiting planets orbiting GJ 9827, a bright (K = 7.2) nearby late K-type dwarf star. GJ 9827 hosts a $1.62\pm0.11$ $R_{\rm \oplus}$ super Earth on a 1.2 day period, a $1.269^{+0.087}_{-0.089}$ $R_{\rm \oplus}$ super Earth on a 3.6 day period, and a $2.07\pm0.14$ $R_{\rm \oplus}$ super Earth on a 6.2 day period.  The radii of the planets transiting GJ 9827 span the transition between predominantly rocky and gaseous planets, and GJ 9827 b and c fall in or close to the known gap in the radius distribution of small planets between these populations. At a distance of 30 parsecs, GJ 9827 is the closest exoplanet host discovered by {\it K2} to date, making these planets well-suited for atmospheric studies with the upcoming {\it James Webb Space Telescope}. The GJ 9827 system provides a valuable opportunity to characterize interior structure and atmospheric properties of coeval planets spanning the rocky to gaseous transition. 

\end{abstract}

\keywords{planetary systems, planets and satellites: detection,  stars: individual (\thisstar)}

\section{Introduction}
With the confirmation of over 3500 planets to date and an additional $\sim$4500 candidates from \Kepler\ \citep{Thompson:2017}, the focus of studying exoplanets has largely shifted from pure discovery to understanding planetary demographics, system architectures, interior structures, and atmospheres. In particular, planets which transit their host stars are valuable for understanding the properties of small planets in detail. Like an eclipsing binary star, combining the transit light curve with radial velocity observations yields a measurement of the mass and radius of a planet relative to its star, which constrain the planet's interior structure. Planetary atmospheres can also be studied if the planet transits. The opacity of a planet's atmosphere depends on its chemical composition and the wavelength of the observation. This causes the apparent size of the planet to change as a function of wavelength. Therefore, by measuring the depth of the transit as a function of wavelength, it is possible to gain insight into the composition and temperature of the planet's atmosphere \citep[this technique is known as transit transmission spectroscopy, ][]{Seager:2000, Brown:2001, Fortney:2003}.

Our ability to study the interior structures and atmospheres of planets, especially small planets ($<$3\rearth) with small radial velocity and atmospheric signals, is highly dependent on the brightness of its host star. The brighter the host star, the easier it is to attain high enough signal-to-noise ratios to search for the small signals produced by small planets. The relative size of the planet to its host star is also highly important for transit transmission spectroscopy. It is easier to detect the small, wavelength-dependent changes in transit depth when planets are larger compared to their host stars, so small stars are more favorable targets than large stars for transit spectroscopy measurements. Therefore, nearby bright small stars with planets are excellent targets for atmospheric characterization \citep{Burrows:2014}. 


Multi-planet systems provide the opportunity to compare the atmospheres and interior structures of different planets while accounting for many confounding variables, like formation history and composition. In some cases, like the recently discovered seven-planet system transiting the nearby late M-dwarf TRAPPIST-1 \citep{Gillon:2016, Gillon:2017}, it is possible to study similarly sized planets across orders of magnitude in incident flux.  In terms of the stellar irradiation of the seven planets, TRAPPIST-1 c resembles Venus, TRAPPIST-1 d resembles the Earth, and TRAPPIST-1 f is similar to Mars \citep{Gillon:2017}. 


However, it would also be desirable to find a multi-planet system suitable for characterization which has planets with different sizes in order to understand the compositions of small planets ranging in size from similar to Earth to about four times the size of Earth. The \Kepler\ mission has found a nearly ubiquitous population of planets with radii larger than the Earth but smaller than Neptune \citep{Batalha:2013,Howard:2012, Petigura:2013, Morton:2014, Christiansen:2015, Dressing:2015}, for which we have no analogue in our own solar system. Recently, the California Kepler Survey (CKS) measured precise radii for over 2000 \Kepler\ planets and found a bimodal distribution in the radii of small planets, with a deficit of planets with radii between 1.5 and 2.0 \rearth, and two peaks in the radius distribution at about 1.3 \rearth\ and 2.5 \rearth \citep{Fulton:2017}. The deficit in radii around 1.5-2 \rearth\ is coincident with the transition \citep{Weiss:2014, Rogers:2015} between predominantly rocky planets (typically smaller than 1.6 \rearth) and planets with substantial gaseous envelopes (typically larger than 1.6 \rearth) as determined from mass measurements of a large number of sub-Neptune-sized planets discovered by \Kepler\ \citep{Wu:2013, Marcy:2014, Hadden:2014, Hadden:2017}. Since most of these planets with mass measurements orbit very close to their host stars (P$<$100 days), they receive a large amount of high-energy irradiation that can evaporate gaseous envelopes made of H/He \citep{Yelle:2004, Tian:2005, Murray-Clay:2009, Owen:2012}. The observed lack of planets with radii of 1.5-2.0 \rearth could be due to these gaseous envelopes being evaporated away and leaving the smaller denser cores \citep{Owen:2017, Jin:2017}. 

\begin{figure*}[!ht]
\vspace{0.3in}
\includegraphics[width=0.99\linewidth, trim = 0 1.5in 0 0]{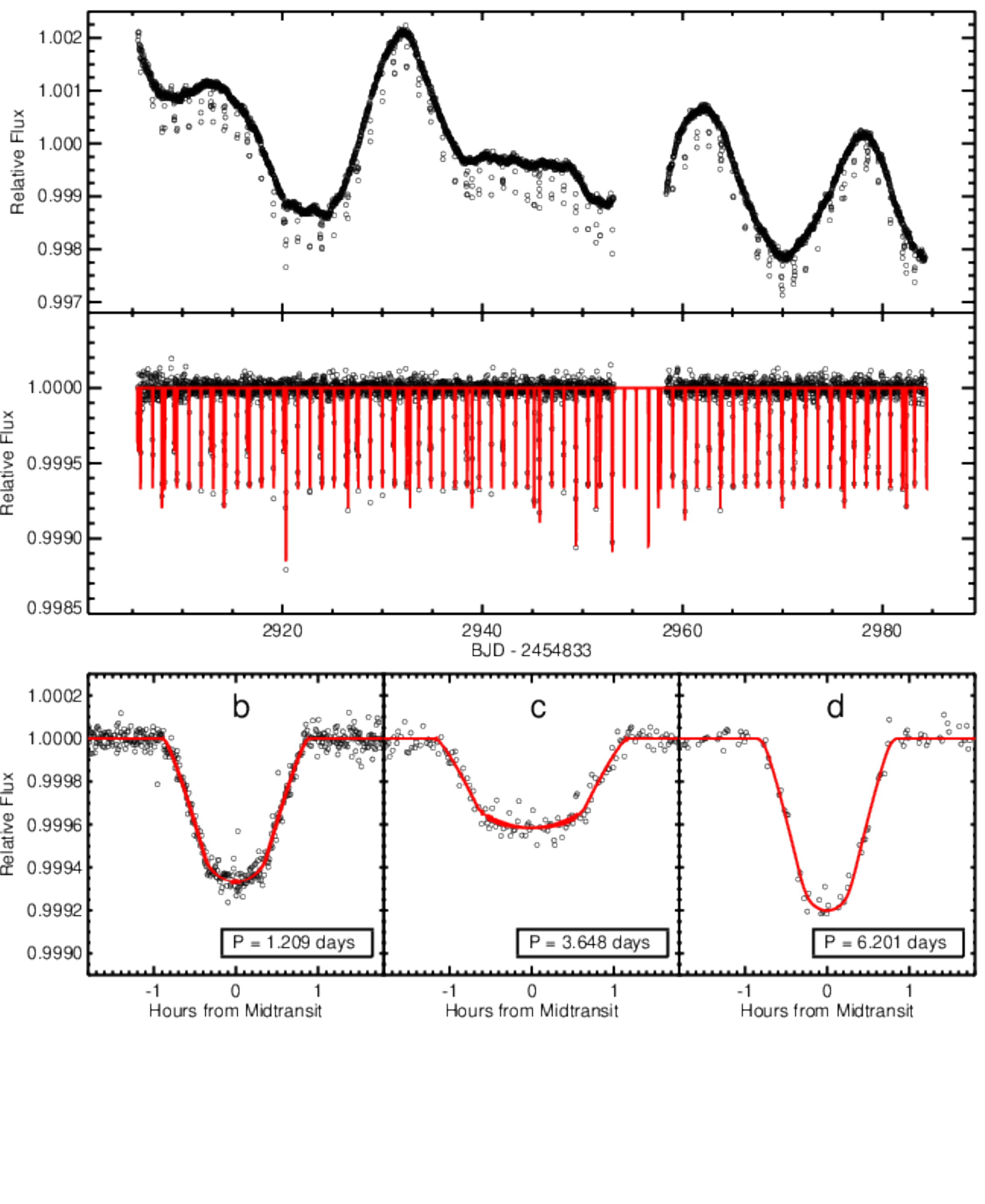}
\caption{(Top) The full K2 light curve of \thisstar from Campaign 12, corrected for systematics using the technique described in \citet{Vanderburg:2014} and \citet{Vanderburg:2016}. (Middle) The corrected K2 lightcurve with best-fit low frequency variability removed. (Bottom) Phase folded \emph{K2} light curves of \thisstar b, c. and d. The observations are plotted in open black circles, and the best fit models are plotted in red.}
\label{figure:LC}
\end{figure*}


In this paper, we present the discovery of three transiting planets orbiting the nearby (d=30.3 $\pm$ 1.6 pc) star \thisstar using data from the {\it K2} mission. The planets transiting \thisstar\ are the closest planets discovered by K2 (surpassing K2-18, at 34$\pm$4 pc \citealt{Montet:2015, Crossfield:2016, Benneke:2017}). \thisstar b, c, and d are all super-Earth sized with radii R$_b$ = \radiusb, R$_c$ = \radiusc, R$_d$ = \radiusd. Planets b ($P_b$ = 1.209d) and c ($P_c$ = 3.648d) orbit about half a percent outside of a 1:3 mean motion resonance, while planet d ($P_d$ = 6.201) orbits far from integer period ratios with the other two planets. The host is a bright (J$\approx$ 8, H $\approx$7.4, K $\approx$ 7.2) nearby late K star, making it an excellent target for atmospheric characterization with the upcoming {\em James Webb Space Telescope} \citep{Gardner:2006}.  The planets span the transition from rocky to gaseous planets, so the characteristics of their atmospheres and interior structures may illuminate how the structure and composition of small planets change with radius.

\begin{deluxetable}{llcc}
\tablecaption{\thisstar\ Magnitudes and Kinematics}
\startdata
\hline
Other identifiers\dotfill & 
       \multicolumn{3}{l}{HIP 115752}\\
	  & \multicolumn{3}{l}{2MASS J23270480-0117108}\\
	  & \multicolumn{3}{l}{EPIC 246389858}\\
\hline
\hline
Parameter & Description & Value & Source\\
\hline
$\alpha_{J2000}$\dotfill	&Right Ascension (RA)\dotfill & 23:27:04.83647			& 1	\\
$\delta_{J2000}$\dotfill	&Declination (Dec)\dotfill & -01:17:10.5816			& 1	\\
\\
B$_T$\dotfill			&Tycho B$_T$ mag.\dotfill & 12.10 $\pm$ 0.178		& 2	\\
V$_T$\dotfill			&Tycho V$_T$ mag.\dotfill & 10.648 $\pm$ 0.069		& 2	\\
$B$\dotfill		& APASS Johnson $B$ mag.\dotfill	& 11.569 $\pm$	0.034		& 3	\\
$V$\dotfill		& APASS Johnson $V$ mag.\dotfill	& 10.250 $\pm$	0.138		& 3	\\
$g'$\dotfill		& APASS Sloan $g'$ mag.\dotfill	& 10.995  $\pm$	0.021		& 3	\\
$r'$\dotfill		& APASS Sloan $r'$ mag.\dotfill	& 	9.845	& 3	\\
$i'$\dotfill		& APASS Sloan $i'$ mag.\dotfill	& 	9.394  & 3	\\
\\
J\dotfill			& 2MASS $J$ mag.\dotfill & 7.984  $\pm$ 0.02		& 4, 5	\\
H\dotfill			& 2MASS $H$ mag.\dotfill & 7.379 $\pm$ 0.04	& 4, 5	\\
K$_S$\dotfill			& 2MASS $K_S$ mag.\dotfill & 7.193 $\pm$ 0.020	& 4, 5	\\
\\
\textit{WISE1}\dotfill		& \textit{WISE1} mag.\dotfill & 6.990 $\pm$ 0.041		& 6	\\
\textit{WISE2}\dotfill		& \textit{WISE2} mag.\dotfill & 7.155 $\pm$ 0.02		& 6 \\
\textit{WISE3}\dotfill		& \textit{WISE3} mag.\dotfill &  7.114 $\pm$ 0.017		& 6	\\
\textit{WISE4}\dotfill		& \textit{WISE4} mag.\dotfill & 6.957 $\pm$0.107		& 6	\\
\\
$\mu_{\alpha}$\dotfill		& NOMAD proper motion\dotfill & 374.4 $\pm$ 2.2 		& 7 \\
                    & \hspace{3pt} in RA (mas yr$^{-1}$)	& & \\
$\mu_{\delta}$\dotfill		& NOMAD proper motion\dotfill 	&  215.7 $\pm$ 1.9 &  7 \\
                    & \hspace{3pt} in DEC (mas yr$^{-1}$) & & \\
\\
$v\sin{i_\star}$\dotfill &  Rotational velocity \hspace{9pt}\dotfill &  1.3$\pm$1.5 \kms & 8 \\
$\mh$\dotfill &   Metallicity \hspace{9pt}\dotfill & -0.5$\pm$0.1 & 8 \\
$\teff$\dotfill &  Effective Temperature \hspace{9pt}\dotfill &  4270$\pm$100 K & 8 \\
log(g)\dotfill &  Surface Gravity \hspace{9pt}\dotfill &  4.9$\pm$0.2 (cgs) & 8 \\
$\pi$\dotfill & Hipparcos Parallax (mas) \dotfill & 32.98 $\pm$ 1.76 & 1 \\
$d$\dotfill & Distance (pc)\dotfill & 30.32$\pm$1.62 & 1 \\
Spec. Type\dotfill & Spectral Type\dotfill & 	K5V & 9 \\

\enddata

\tablecomments{
References are: $^1$\citet{vanLeeuwen:2007},$^2$\citet{Hog:2000} $^3$\citet{Henden:2015},$^4$\citet{Cutri:2003}, $^5$\citet{Skrutskie:2006}, $^6$\citet{Cutri:2014}, $^7$\citet{Zacharias:2004}, $^8$\citet{Houdebine:2016}, $^9$\citet{Reid:1995}
}

\label{tab:LitProps}
\end{deluxetable}

\section{Observations and Archival Data}
\subsection{K2 Photometry}
In May 2013 the \Kepler\ spacecraft experienced a failure of the second of its four reaction wheels, ending its primary mission. However, the \Kepler\ spacecraft has been re-purposed to obtain high precision photometry for $\sim$80 days at a time on a set of fields near the ecliptic in its extended {\it K2} mission \citep{Howell:2014}. During K2 Campaign 12, \thisstar was observed from UT 2016 December 16 until UT 2017 March 04. We identified  \thisstar as a candidate planet host after downloading all of the \Kepler-pipeline calibrated target pixel files from the Mikulski Archive for Space Telescopes, producing light curves, and correcting for K2 spacecraft systematics following \citet{Vanderburg:2014} and \citet{Vanderburg:2016}. We then searched the resulting light curves for transiting planet candidates using the pipeline described by \citet{Vanderburg:2016}. Among the objects uncovered in our search were three super-Earth-sized planet candidates with periods of 1.2, 3.6, and 6.2 days around the nearby star \thisstar. After we identified the signals, we re-processed the K2 light curve to simultaneously fit the transits, stellar variability, and K2 systematics. We flattened the light curve by dividing away the best-fit stellar variability (which we modeled as a basis spline with breakpoints every 0.75 days) from our simultaneous fit to the light curve. The final lightcurve has a noise level of 39 ppm per 30 minute cadence exposure, and a 6 hour photometric precision of 9 ppm. See Figure \ref{figure:LC} for the final light curve.

The K2 light curve shows rotational stellar variability on \thisstar\ with a typical amplitude of about 0.2\% peak to peak (See Figure \ref{figure:LC}). We calculated the autocorrelation function of the K2 light curve, and find a rotation period of 31 $\pm$ 1 days, although it is possible the true rotation period is at about 16 days, or half our best estimate. The autocorrelation function preferred a 31 day period most likely because of the flatness at BJD$_{\rm TDB}$-2454833 = 2945 instead of another peak.

\begin{figure*}[!ht]
\vspace{0.3in}
\centering
\includegraphics[width =1.0\linewidth,angle =0]{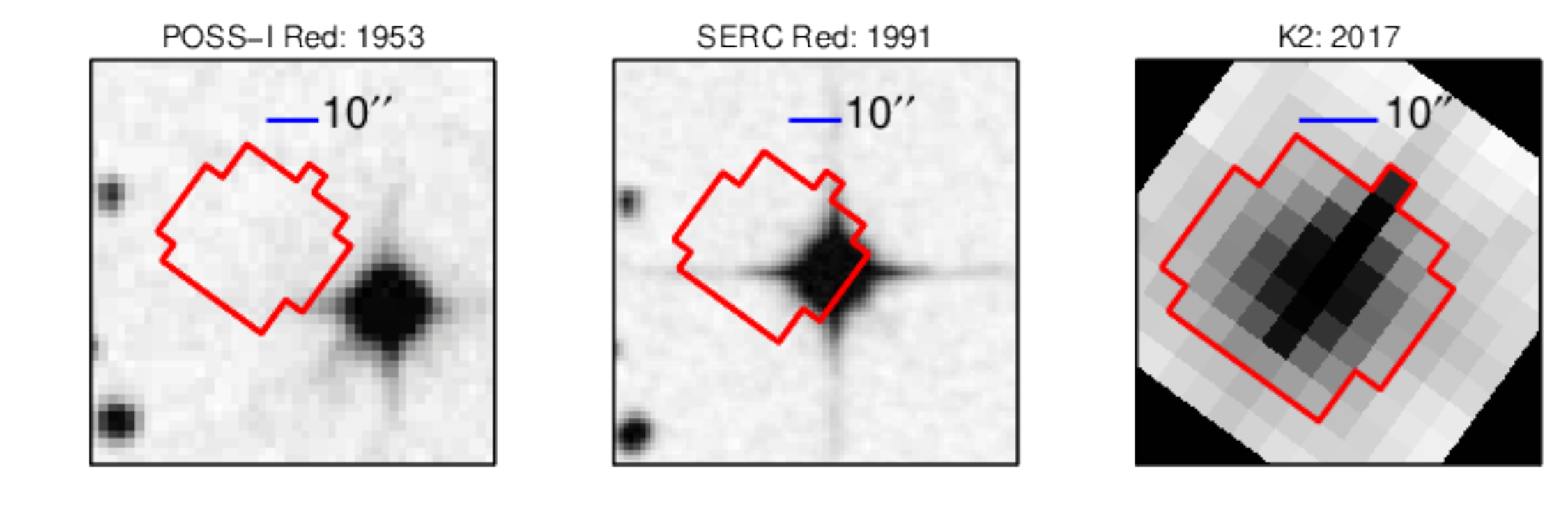}
\vspace{-0.35in}

\includegraphics[width=0.5\linewidth, angle =90, trim = 0 0in 0 0]{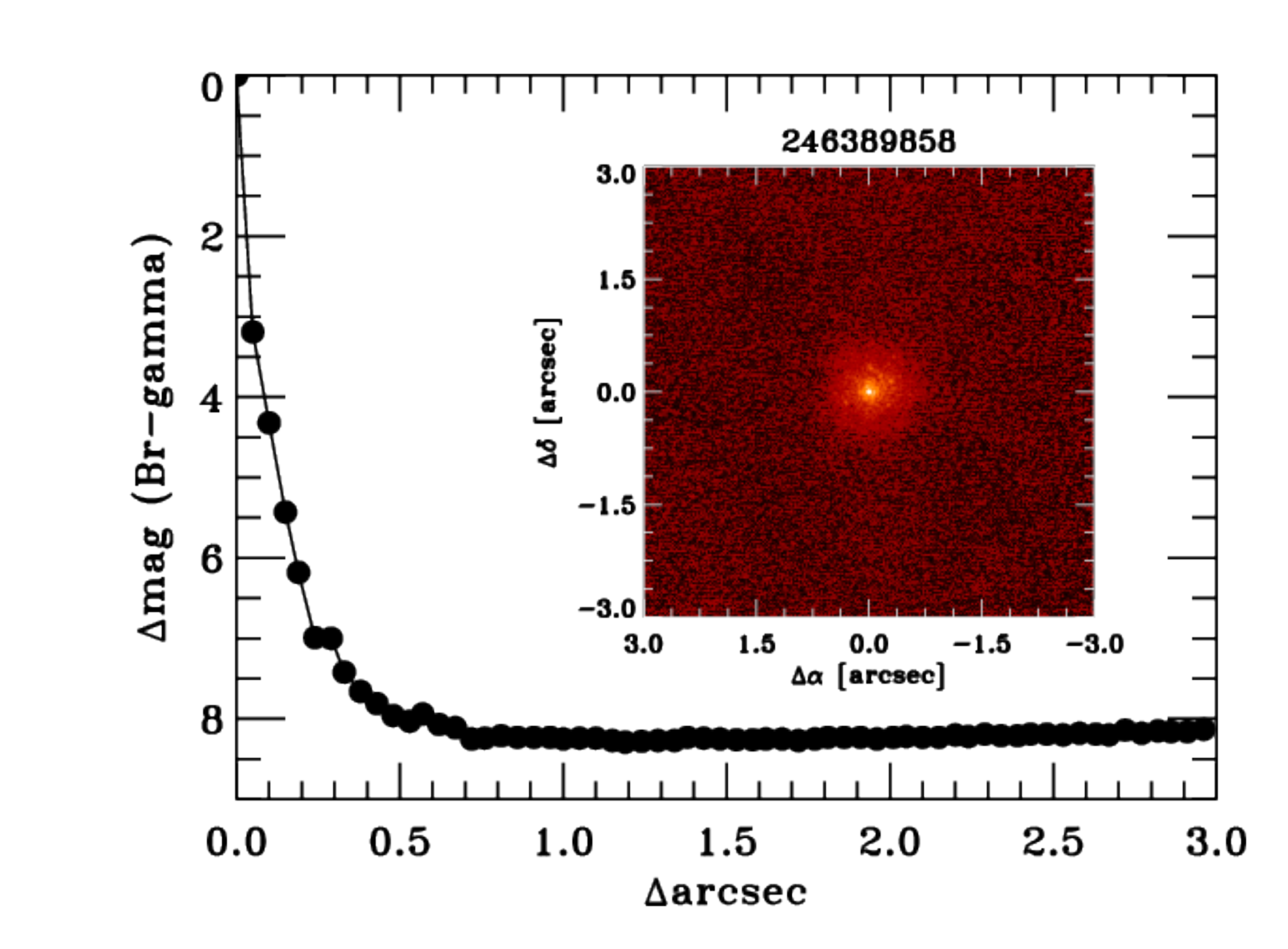}
\caption{(Top Left) Archival imaging from the National Geographic Society Palomar Observatory Sky Survey (NGS POSS) of \thisstar\ taken with a red emulsion in 1953.  (Top Middle) Archival imaging from the ESO/SERC survey of \thisstar\ taken with a red emulsion in 1991.  (Top Right) Summed image of \thisstar\ from K2 observations. The aperture selection is described in \citet{Vanderburg:2016b}. (Bottom) The Keck Br-$\gamma$ contrast curve and image (inset) of \thisstar. We find no evidence of any additional components in the system. }
\label{figure:AO}
\end{figure*}

\subsection{Archival Spectroscopy}

As part of a survey of nearby Solar-type stars, \thisstar was observed on UT 2000 Aug 31 using the Center for Astrophysics (CfA) Digital Speedometer on the 1.5~m Wyeth Reflector at the Oak Ridge Observatory in the town of Harvard, Massachusetts. The Digital Speedometer measured an absolute RV of 31.2 \kms with an approximate accuracy of $\sim$0.3 \kms (Latham, private communication). \thisstar was also observed on UT 2010 Oct 08 and UT 2011 Aug 06 using a CORAVEL-type spectrometer at Vilnius University Observatory, which measured absolute RVs of \thisstar on these dates of 32.6 \kms and 31.1 \kms, respectively \citep{Sperauskas:2016}. Using the equations given in \citet{Johnson:1987}, the UVW space velocities of \thisstar were estimated to be (U,V,W) = ($\minus$59.2, 20.9, 30.6) \kms \citep{Sperauskas:2016}. Using the probability distributions of \citet{Reddy:2006}, \thisstar is predicted to be a member of the Galactic thin disk. From these observations, we see no evidence of any large RV variation over the span of over 10 years.

\thisstar\ was also observed twice in 2004 with the High Accuracy Radial Velocity Planet Searcher (HARPS) spectrograph as part of the guaranteed time collaboration's planet search, but not enough observations were taken to identify the small planet candidates we find. Later, \citet{Houdebine:2016} used a principal component analysis based method to analyze the HARPS spectra and estimate stellar parameters. They found: \teff = 4270$\pm$100, \feh = -0.5$\pm$0.1 dex, \logg = 4.9$\pm$0.2, and $v\sin{i_\star}$ = 1.3$^{+1.5}_{-1.3}$ \kms. From the Hipparcos parallax and an analysis of the spectral energy distribution (SED), \citet{Houdebine:2016} estimated the radius of \thisstar to be \rstar \ of 0.623$\pm$0.082 \rsun. In this paper, we adopt the spectroscopic parameters from \citet{Houdebine:2016} but derive our own stellar mass and radius for our global modeling (described in Section \ref{sec:GlobalModel}). 

\subsection{Archival Seeing-Limited Imaging}
Using archival observations from the National Geographic Society Palomar Observatory Sky Survey (NGS POSS) from 1953 and 1991 (ESO/SERC), we looked for nearby bright companions that may dilute our observed transit depths. \thisstar has a high proper motion ($\mu_{\alpha}$ = 374.4 mas and $\mu_{\delta}$ = 215.7 mas), and has moved nearly 30$\arcsec$ from its original position when the POSS image was taken in 1953 (See Figure \ref{figure:AO}). In 1953, \thisstar\ was outside of the region of sky enclosed within the photometric aperture we use to produce its modern K2 light curve. No background stars are present inside our K2 photometric aperture down to the POSS limiting magnitude of about R = 20, a full 10 magnitudes fainter than \thisstar. Since all three transit signals around \thisstar\ have depths greater than 100 ppm, the maximum depth of a transit caused by a background star 10 magnitudes fainter than \thisstar, we can use ``patient imaging'' to confidently rule out background stars as the sources of these transit signals.  

\subsection{Keck/NIRC2 AO Imaging}
Using the Near Infrared Camera 2 (NIRC2) behind the natural guide star adaptive optics system at the W. M. Keck Observatory, we obtained high resolution images of \thisstar\ using the Br-$\gamma$ filter on UT 2017 August 19. NIRC2 has a 1024$\times$1024 pixel array with a 9.942 mas pix$^{-1}$ pixel scale. The lower left quadrant of the NIRC2 array suffers from a higher noise level and a 3-point dither pattern was adopted excluding this regime of the detector. After flat-fielding and sky subtraction, each observation was shifted and co-added, resulting in the final image shown in Figure \ref{figure:AO}. No other star was detected in the 10$\arcsec$ field-of-view. To determine our sensitivity to companions, we inject simulated sources into the final image that have a signal to noise of 5. Figure \ref{figure:AO} shows the 5$\sigma$ sensitivity as a function of spatial separation from \thisstar, and the inset shows the image itself.

\begin{center}
\begin{figure}[!ht]
\centering
\includegraphics[width=0.9\linewidth, angle = 0,trim = 0in 0.5in 0in 0in]{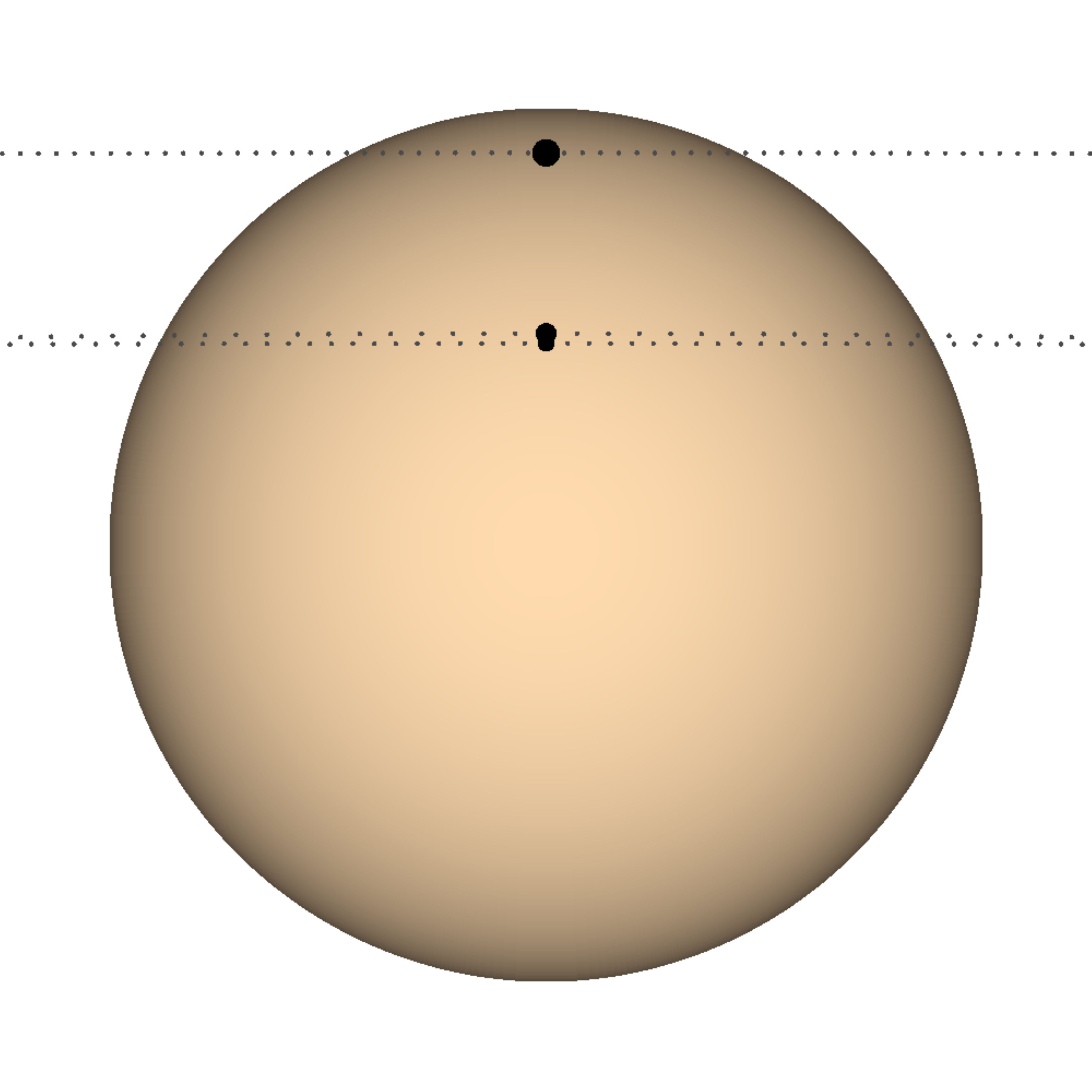}
\caption{A diagram of the \thisstar system geometry shown with all planets at their respective transit centers. From top to bottom, the planets are d, b, and c. The color of the star matches its effective temperature, the planets are to scale with respect to each other and the host star, and the limb darkening matches our best-fit model in the Kepler band. The grey dots trace the orbital path of the planet, with a dot every three minutes. The curvature of planet b's orbit is plainly visible. $\Omega$ for each planet (a rotation of the path about the center of the star) is assumed to be zero. Note the mutual inclinations may be much larger than implied here due to the ambiguity between the inclination and 180 degrees minus the inclination. Also note that, while this is the most likely model, the uncertainty in the impact parameters for planets b and c allow them to be non-overlapping (see Figure \ref{figure:simultaneous}).}
\label{figure:chords}
\end{figure}
\end{center}

\begin{figure*}[!ht]
\vspace{0.3in}
\includegraphics[width=0.49\linewidth, trim = 0 5.5in 0 0]{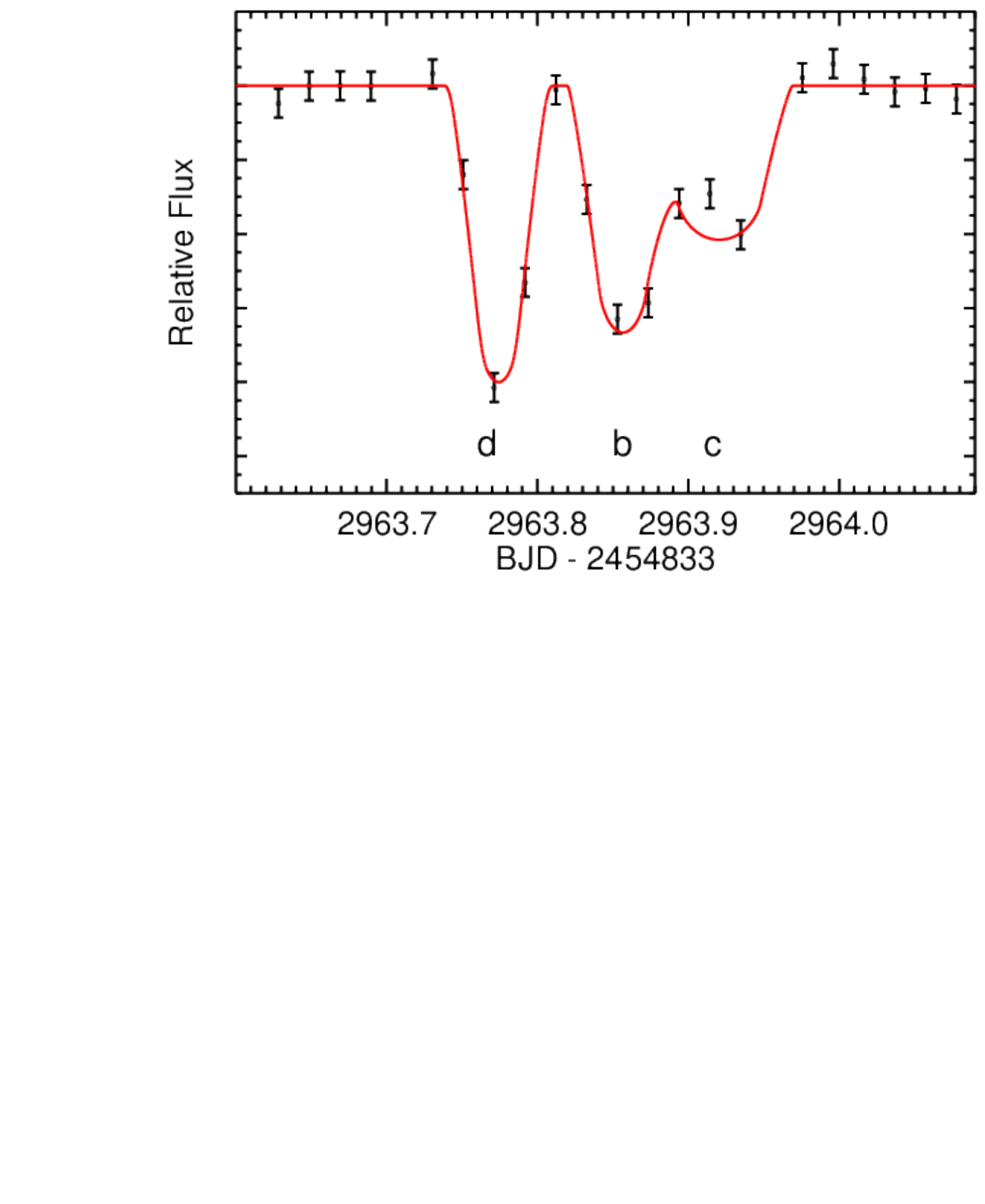}\includegraphics[width=0.49\linewidth, trim = 0 0 0 0]{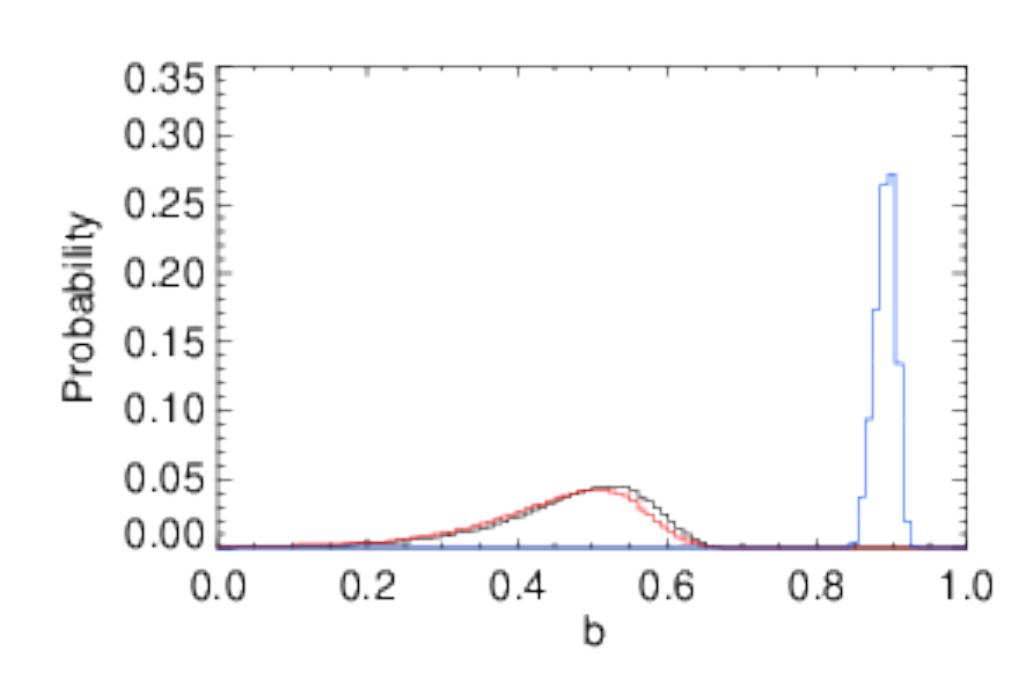}
\caption{(Left) The corrected K2 lightcurve for \thisstar showing a simultaneous transit of b and c with the EXOFASTv2 model shown in red. (Right) The probability distribution of the impact parameter for \thisstar b (black), c (red), and d (blue). We cannot rule out the possibility of mutual transits of \thisstar b and c.}
\label{figure:simultaneous}
\end{figure*}

\section{System Modeling} 
\label{sec:GlobalModel}

Making use of the flattened {\it K2} lightcurves, the Hipparcos parallax, and stellar parameters, we perform a global fit of the \thisstar system using EXOFASTv2 \citep[][Eastman et. al., in prep]{Eastman:2013,Eastman:2017}. EXOFASTv2 is based heavily on EXOFAST, but a large fraction of the code has been rewritten to be more flexible. EXOFASTv2 can now, among other things, simultaneously fit multiple planets, incorporate characterization observations (like Doppler Tomography), and simultaneously perform an SED within the global fit. EXOFASTv2 has a few major conceptual changes. First, an the error scaling term for the transit photometry is now fit within the Markov chain Monte Carlo (MCMC). Also, the fit uses the stepping parameters log(\mstar) and age instead of \ar and \logg. EXOFASTv2 has previously been used to determine parameters for the HD~106315 system \citep{Rodriguez:2017b}. 

Because \thisstar is relatively low-mass with marginal applicability to both the Torres relations \citep{Torres:2010} and YY isochrones \citep{Yi:2001}, we disable those constraints within the global model. To determine the mass and radius of \thisstar, we interpolated the absolute $K_S$-band magnitude onto a grid of stellar evolutionary models and the semi-empirical $M_K$-$M_*$ and $M_K$-$R_*$ relations from \citet{Mann:2015b}. We assumed a main-sequence but unknown age (0.5-10\,Gyr), a metallicity of -0.15$\pm$0.2, and a solar [$\alpha$/Fe]. This metallicity is based on the star's color-magnitude position \citep{Neves:2012} and $JHK$ colors \citep{Mann:2013a, Newton:2014}. However, the $K$-band is selected specifically because it shows a weak dependence with $M_*$ and $R_*$, so adopting a lower metallicity as found by our spectral fitting does not significantly change the result. We tried both the Mesa Isochrones and Stellar Tracks \citep[MIST,][]{MIST0,MIST1} and Dartmouth Stellar Evolution Program \citep[DSEP,][]{Dotter:2008} models, yielding radii of $0.60\pm0.02R_*$ and $0.59\pm0.03R_*$ and masses of $0.63\pm0.03M_*$ and $0.61\pm0.04M_*$ respectively. The relations from \citet{Mann:2015}, which are anchored in radii from long-baseline optical interferometry \citep{Boyajian:2012}, produced a radius estimate of $0.64\pm0.03R_*$ and mass of $0.66\pm0.02M_*$. Errors account for uncertainties in the parallax and $K_S$-band magnitude. 

\thisstar lands in a region of parameter space where weak molecular bands can form, where models are known to systematically underestimate the radii. However it also lands at the bright limit of the \citet{Mann:2015b} relations, around which the calibration stars are preferentially metal-rich when compared to GJ9827 (which would lead to an overestimated radius). Instead, we adopt more conservative parameters of 0.63$\pm$0.03 \mstar and 0.61$\pm$0.03R \rstar for \thisstar, which encompasses all values above with comparable uncertainties. These values for \rstar and \mstar were used as priors for the global fit.



We performed a separate global fit using the broad band photometry summarized in Table \ref{tab:LitProps}, the Hipparcos parallax, an upper limit on extinction from \citet{Schlegel:1998}, to derive the radius of the star. This fit recovered a consistent stellar radius and uncertainty to \citet{Houdebine:2016}, but the stellar metalicity was driven too high, perhaps biased unfairly by the lack of SED models for such metal-poor stars. 

From the \citet{Houdebine:2016} analysis of HARPS South spectra combined with an SED analysis of \thisstar, we set a prior on \teff of 4270$\pm$100 K. Additionally, we imposed a prior on the parallax from Hipparcos \citep{vanLeeuwen:2007} Such a metal poor, low-mass star may suffer from systematic biases in the limb darkening and gaps in the parameter tables. While a small error in the limb darkening is well within the uncertainty of the K2 lightcurve, allowing it to be fit within the the global fit may work backward to bias the \logg, \teff, and \feh \ from which they are derived.  Therefore, while the limb darkening values can be derived within EXOFASTv2 using the \citet{Claret:2011}, we place a uniform prior of $\mu$1 = 0.44 $\pm$ 0.1 and $\mu$2 = 0.26 $\pm$ 0.1. The starting values were determined using the EXOFAST online tool\footnote{http://astroutils.astronomy.ohio-state.edu/exofast/limbdark.shtml} \citet{Eastman:2016}.


The system parameters determined from our global fit are shown in Table \ref{tbl:GJ9827} and a diagram of the system geometry is shown in Figure \ref{figure:chords}. 
\begin{deluxetable*}{lcccc}
\tablecaption{Median values and 68\% confidence interval for GJ~9827.}
\tablehead{\colhead{~~~Parameter} & \colhead{Units} & \multicolumn{3}{c}{Values}}
\centering
\startdata
\multicolumn{2}{l}{Stellar Parameters}\\
\hline\\
~~~~$M_*$\dotfill &Mass (\msun)\dotfill &$0.614^{+0.030}_{-0.029}$\\
~~~~$R_*$\dotfill &Radius (\rsun)\dotfill &$0.613^{+0.033}_{-0.034}$\\
~~~~$\rho_*$\dotfill &Density (cgs)\dotfill &$3.76^{+0.75}_{-0.57}$\\
~~~~$\log{g}$\dotfill &Surface gravity (cgs)\dotfill &$4.651^{+0.055}_{-0.050}$\\
~~~~$T_{eff}$\dotfill &Effective Temperature (K)\dotfill &$4269^{+98}_{-99}$\\
\\\hline
\multicolumn{2}{l}{Planetary Parameters:}&b &c&d\\
\hline\\
~~~~$a$\dotfill &Semi-major axis (AU)\dotfill &$0.01888^{+0.00030}_{-0.00031}$&$0.03942^{+0.00062}_{-0.00064}$&$0.05615^{+0.00089}_{-0.00091}$\\
~~~~$P$\dotfill &Period (days)\dotfill &$1.2089802^{+0.0000084}_{-0.0000081}$&$3.648083^{+0.000060}_{-0.000058}$&$6.201467^{+0.000062}_{-0.000061}$\\
~~~~$M_P$\dotfill &Mass (\me)\dotfill &$3.42^{+1.2}_{-0.76}$&$2.42^{+0.75}_{-0.49}$&$5.2^{+1.8}_{-1.2}$\\
~~~~$R_P$\dotfill &Radius (\re)\dotfill &$1.62\pm0.11$&$1.269^{+0.087}_{-0.089}$&$2.07\pm0.14$\\
~~~~$i$\dotfill &Inclination (Degrees)\dotfill &$85.73^{+1.2}_{-0.96}$&$88.05^{+0.64}_{-0.48}$&$87.39^{+0.20}_{-0.18}$\\
~~~~$\rho_P$\dotfill &Density (cgs)\dotfill &$4.50^{+1.5}_{-0.98}$&$6.4^{+2.0}_{-1.1}$&$3.23^{+1.1}_{-0.72}$\\
~~~~$logg_P$\dotfill &Surface gravity \dotfill &$3.110^{+0.12}_{-0.098}$&$3.163^{+0.11}_{-0.082}$&$3.07^{+0.12}_{-0.10}$\\
~~~~$T_{eq}$\dotfill &Equilibrium temperature (K)\dotfill &$1172\pm43$&$811\pm30$&$680\pm25$\\
~~~~$\Theta$\dotfill &Safronov Number \dotfill &$0.00460^{+0.0015}_{-0.00096}$&$0.0086^{+0.0025}_{-0.0016}$&$0.0162^{+0.0052}_{-0.0035}$\\
~~~~$\fave$\dotfill &Incident Flux (\fluxcgs)\dotfill &$0.429^{+0.066}_{-0.060}$&$0.098^{+0.015}_{-0.014}$&$0.0485^{+0.0074}_{-0.0068}$\\
~~~~$T_C$\dotfill &Time of Transit (\bjdtdb)\dotfill &$2457738.82588^{+0.00030}_{-0.00031}$&$2457742.19944^{+0.00063}_{-0.00068}$&$2457740.96111\pm0.00044$\\
~~~~$T_P$\dotfill &Time of Periastron (\bjdtdb)\dotfill &$2457738.82588^{+0.00030}_{-0.00031}$&$2457742.19944^{+0.00063}_{-0.00068}$&$2457740.96111\pm0.00044$\\
~~~~$T_S$\dotfill &Time of eclipse (\bjdtdb)\dotfill &$2457739.43037\pm0.00030$&$2457744.02348^{+0.00061}_{-0.00066}$&$2457744.06185\pm0.00041$\\
~~~~$T_A$\dotfill &Time of Ascending Node (\bjdtdb)\dotfill &$2457738.52363^{+0.00030}_{-0.00031}$&$2457741.28742^{+0.00065}_{-0.00069}$&$2457739.41074\pm0.00045$\\
~~~~$T_D$\dotfill &Time of Descending Node (\bjdtdb)\dotfill &$2457739.12812^{+0.00030}_{-0.00031}$&$2457743.11146^{+0.00062}_{-0.00067}$&$2457742.51148^{+0.00042}_{-0.00043}$\\
~~~~$K$\dotfill &RV semi-amplitude (m/s)\dotfill &$2.84^{+0.97}_{-0.64}$&$1.39^{+0.44}_{-0.29}$&$2.50^{+0.86}_{-0.58}$\\
~~~~$logK$\dotfill &Log of RV semi-amplitude \dotfill &$0.45^{+0.13}_{-0.11}$&$0.14^{+0.12}_{-0.10}$&$0.40^{+0.13}_{-0.12}$\\
~~~~$M_P\sin i$\dotfill &Minimum mass (\me)\dotfill &$3.41^{+1.2}_{-0.76}$&$2.42^{+0.75}_{-0.49}$&$5.2^{+1.8}_{-1.2}$\\
~~~~$M_P/M_*$\dotfill &Mass ratio \dotfill &$0.0000168^{+0.0000058}_{-0.0000038}$&$0.0000119^{+0.0000037}_{-0.0000025}$&$0.0000254^{+0.0000089}_{-0.0000060}$\\
~~~~$R_P/R_*$\dotfill &Radius of planet in stellar radii \dotfill &$0.02420^{+0.00040}_{-0.00047}$&$0.01899^{+0.00034}_{-0.00037}$&$0.03093^{+0.00065}_{-0.00059}$\\
~~~~$a/R_*$\dotfill &Semi-major axis in stellar radii \dotfill &$6.62^{+0.41}_{-0.35}$&$13.83^{+0.86}_{-0.74}$&$19.7^{+1.2}_{-1.0}$\\
~~~~$d/R_*$\dotfill &Separation at mid transit \dotfill &$6.62^{+0.41}_{-0.35}$&$13.83^{+0.86}_{-0.74}$&$19.7^{+1.2}_{-1.0}$\\
~~~~$b$\dotfill &Impact parameter \dotfill &$0.493^{+0.080}_{-0.12}$&$0.469^{+0.085}_{-0.13}$&$0.896^{+0.012}_{-0.016}$\\
~~~~$\delta$\dotfill &Transit depth \dotfill &$0.000586^{+0.000019}_{-0.000023}$&$0.000361^{+0.000013}_{-0.000014}$&$0.000957^{+0.000041}_{-0.000036}$\\
~~~~$Depth$\dotfill &Flux decrement at mid transit \dotfill &$0.000586^{+0.000019}_{-0.000023}$&$0.000361^{+0.000013}_{-0.000014}$&$0.000957^{+0.000041}_{-0.000036}$\\
~~~~$P_T$\dotfill &A priori non-grazing transit prob \dotfill &$0.1474^{+0.0082}_{-0.0086}$&$0.0710^{+0.0040}_{-0.0042}$&$0.0492^{+0.0027}_{-0.0029}$\\
~~~~$P_{T,G}$\dotfill &A priori transit prob \dotfill &$0.1547^{+0.0088}_{-0.0092}$&$0.0737^{+0.0042}_{-0.0044}$&$0.0524^{+0.0030}_{-0.0031}$\\
~~~~$T_{FWHM}$\dotfill &FWHM duration (days)\dotfill &$0.05083^{+0.00080}_{-0.00072}$&$0.0743^{+0.0011}_{-0.0012}$&$0.04398^{+0.0010}_{-0.00094}$\\
~~~~$\tau$\dotfill &Ingress/egress duration (days)\dotfill &$0.00163^{+0.00023}_{-0.00022}$&$0.00181^{+0.00025}_{-0.00024}$&$0.00708^{+0.00098}_{-0.00094}$\\
~~~~$T_{14}$\dotfill &Total duration (days)\dotfill &$0.05249^{+0.00074}_{-0.00071}$&$0.0761\pm0.0011$&$0.0511^{+0.0011}_{-0.0010}$\\
\hline
\multicolumn{2}{l}{Wavelength Parameters:}& Kepler &&\\
\hline\\
~~~~$u_{1,Kepler}$\dotfill &linear limb-darkening coeff \dotfill &$0.417^{+0.069}_{-0.053}$\\
~~~~$u_{2,Kepler}$\dotfill &quadratic limb-darkening coeff \dotfill &$0.240^{+0.075}_{-0.059}$\\
\hline
\multicolumn{2}{l}{Transit Parameters:}& Kepler &&\\
\hline\\
~~~~$\sigma^{2}$\dotfill &Added Variance \dotfill &$-0.000000000002^{+0.000000000037}_{-0.000000000036}$\\
~~~~$F_0$\dotfill &Baseline flux \dotfill &$0.99999999^{+0.00000069}_{-0.00000070}$\\
\enddata
\label{tbl:GJ9827}
\end{deluxetable*}
\vspace{0.5in}

\section{Statistical Validation}
To validate the planetary nature of the candidates identified to be transiting \thisstar, we use the statistical techniques of \citet{morton2012} implemented in the \texttt{vespa} software package \citep{morton2015}. Using the location of the system in the sky and observational constraints, \texttt{vespa} calculates the astrophysical false positive probability (FPP) of the transiting planet candidates. This takes into account the possibility of hierarchical companions or background objects that could lead to a false identification of a transiting planet. Since \thisstar hosts multiple planets it is very unlikely that all three planet candidates are false positives\footnote{However, the chance that one of them is a false positive is harder to rule out \citep{Latham:2011}.}.  Previous works have calculated a ``multiplicity boost'' that reduces the false positive probability for multi-planet systems transiting a star in the original \Kepler\ and {\it K2} fields \citep{lissauer:2012, Sinukoff:2016, Vanderburg:2016c}. After applying the multiplicity boost to the \texttt{vespa} determined FPP for the planets transiting \thisstar, we estimate a FPP of $2\times10^{-6}$, $6\times10^{-7}$, and $6\times10^{-10}$ for b, c, and d. Therefore, \thisstar b, c, and d are validated exoplanets.

\begin{table*}
 \centering
 \caption{The Best Confirmed Planets for Transmission Spectroscopy with R$_P$ $<$ 3\rearth}
 \label{tbl:S/N}
 \begin{tabular}{ccccc}
    \hline
    \hline
    Planet & R$_P$(\rearth) &  S/N$^a$ & Reference \\
    \hline
 GJ 1214 b  & 2.85$\pm$0.20 & 1.00 & \citet{Charbonneau:2009}\\
 55 Cnc e$^{b}$  & 1.91$\pm$0.08 & 0.41 & \citet{Dawson:2010}\\
 HD 97658 b  & 2.34$^{+0.17}_{-0.15}$ & 0.36 &\citet{Dragomir:2013}\\
 TRAPPIST-1f & 1.045$\pm$0.038 & 0.24 & \citep{Gillon:2017}\\ 
 \textbf{\thisstar b }& \radiusb  & \textbf{0.14}&  \textbf{this work} \\
 HD 3167 c & 2.85$^{+0.24}_{-0.15}$ & 0.14 & \citet{Vanderburg:2016c, Christiansen:2017}\\
 HIP 41378 b  & 2.90$\pm$0.44 & 0.14 &\citet{Vanderburg:2016b} \\
 \textbf{\thisstar d} & \radiusd & \textbf{0.13} &  \textbf{this work} \\
 K2-28 b  & 2.32$\pm$0.24 & 0.12 &  \citet{Hirano:2016} \\
 HD 106315 b & 2.5$\pm$0.1 & 0.10 &\citep{Crossfield:2017, Rodriguez:2017b} \\
   \hline
    \hline
 \end{tabular}
\begin{flushleft} 
 \footnotesize{ \textbf{\textsc{NOTES:}}
$^a$The predicted signal-to-noise ratios relative to GJ 1214 b. All values used in determining the signal-to-noise were obtained from the NASA Exoplanet Archive \citep{Akeson:2013}. If a system did not have a reported mass on NASA Exoplanet Archive or it was not a 2$\sigma$ result, we used the \citet{Weiss:2014} Mass-Radius relationship to estimate the planet's mass. $^b$Our calculation for the S/N of 55 Cnc e assumes a H/He envelope since it falls just above the pure rock line determined by \citet{Zeng:2016}. However, 55 Cnc e is in a ultra short period orbit, making it unlikely that it would hold onto a thick H/He envelope.\\
}
\end{flushleft}
\end{table*}

\section{Discussion}
The proximity of \thisstar and its planetary architecture make it a compelling system worth further characterization. At $\approx$30 parsecs, this is the closest exoplanet system discovered by {\it K2} to date and one of the few stars to have multiple transiting terrestrial sized exoplanets that are well-suited for both mass measurements and atmospheric characterization. The host star is quite bright (V=10.3, J=8) and the measured planet radii of \thisstar b, c, and d are \radiusb, \radiusc, and \radiusd. As mentioned before, there is a known dichotomy in the sizes of short period ($<$100 days) small planets where planets are more commonly found to be less than 1.5\rearth or larger than 2.0\rearth \citep{Fulton:2017}. Based on the mass measurements of planets in these two regimes, the larger planets are less dense and consistent with having a H/He envelope. It is thought that planets smaller than $\sim$1.6 \rearth have lost this outer H/He envelope leaving the rocky core, explaining their higher densities and a lack of planets with radii of 1.5 to 2.0 \rearth \citep{Weiss:2014, Rogers:2015}. The three known planets orbiting \thisstar provide a rare opportunity to perform a comparative study since \thisstar c is $<$1.5\rearth, \thisstar d is $>$ 2.0 \rearth, and \thisstar b lands right in this deficiency gap. This system may shed light on the evolution of planets within this radius regime. 

Using the \citet{Weiss:2014} mass-radius relations, we estimate the mass of \thisstar b, c, and d to be $4.26^{+0.54}_{-0.49}$ \mearth, $2.63^{+1.59}_{-1.00}$ \mearth, and $5.32^{+0.68}_{-0.62}$ \mearth. Within our global model, EXOFASTv2 estimated the masses using the \citet{Chen:2017} mass-radius relations to be $3.52^{+1.4}_{-0.93}$ \mearth, $2.46^{+0.89}_{-0.75}$ \mearth, and $5.2^{+2.1}_{-1.5}$ \mearth. We note that the \citet{Weiss:2014} planet mass uncertainties ignore any uncertainty in the mass--radius relation itself, and is only due to the uncertainty in the radius.
The \citet{Chen:2017} estimated masses correspond to RV semi-amplitudes of $2.34^{+0.90}_{-0.54}$ \ms, $1.08^{+0.44}_{-0.25}$ \ms, and $2.01^{+0.79}_{-0.48}$ \ms. \citet{Houdebine:2016} measured the rotational velocity of \thisstar to be $<$2 \kms, making the planets around \thisstar well-suited for precise RV observations with current spectroscopic facilities to measure their masses. The rotation period of \thisstar\ is either 31d or 16d, well separated from the orbital periods of the planets, so it should be possible to filter away signals from stellar activity using techniques like Gaussian process regression \citep{Haywood:2014}.

To better understand the feasibility of characterizing the atmospheres of the three planets orbiting \thisstar, we calculate the atmospheric scale height and an expected signal-to-noise per transit following the description given in \citet{Vanderburg:2016c}. We repeat this calculation for all known planets where $R_p$ $<$ 3\rearth using NASA's Exoplanet archive \citep{Akeson:2013}. It is expected that both \thisstar\ b and d might have thick gaseous atmospheres \citep{Weiss:2014}, while \thisstar\ d likely does not have a thick envelope. We find that \thisstar\ b and d are two of the best small ($R_p$ $<$ 3\rearth) exoplanets for detailed atmospheric characterization (See Table \ref{tbl:S/N})\footnote{We note that signal to noise is not everything. This calculation makes no assumptions about clouds or the presence of high-mean molecular weight atmospheres. The potential pitfalls of making these assumptions are illustrated by GJ 1214, which according to our calculation is the most amenable small planet to atmospheric characterization, but which shows no atmospheric features, likely due to the presence of clouds, hazes, or aerosols \citep{Kreidberg:2014}.}. By studying their atmospheric compositions, we may better understand the observed dichotomy in planetary composition observed at $\sim$1.6 \rearth. All calculations are done using the H-band magnitude of the stars to test the feasibility of characterizing the planet's atmosphere with the {\it Hubble Space Telescope's} Wide Field Camera 3 instrument and the upcoming suite of instruments that will be available on the James Webb Space Telescope. At a J-band magnitude of 8, \thisstar is near the expected saturation limit of the JWST instruments but should be accessible to all four instrument suites allowing for a high S/N with a relatively short exposure time: Near Infared Camera (NIRCam), Near Infrared Imager and Slitless Spectrograph (NIRISS), Near-Infrared Spectrograph (NIRSpec), and the Mid-Infrared Instrument (MIRI) \citep{Beichman:2014}. The brightness of the \thisstar system makes it a great target for NIRCam's Dispersed Hartmann Sensor (DHS) mode \citep{Schlawin:2017}.

The short orbital periods of the three \thisstar\ planets and the near 1 to 3 period commensurability between \thisstar\ b and c provides opportunities to observe overlapping transits of the three planets, as shown in Figure \ref{figure:simultaneous}. The simultaneous transit on UT 2017 Feb 11 of \thisstar b and c shows one discrepant datapoint which misses the EXOFASTv2 model. This kind of discrepancy might be explained by a mutual transit, where \thisstar c actually transits both \thisstar and planet b simultaneously, which is not modeled by EXOFASTv2. However, at the observed time of this observation, the transit of planet b likely would have already completed (unless there was a significant transit timing variation). We do not find any convincing evidence of mutual transits in our analysis but based on the probability of each planet's impact parameters (See Figure \ref{figure:simultaneous}), we are not able to rule out this possibility.

\section{Conclusion}
We present the discovery of three transiting planets orbiting the nearby late K-type star, \thisstar. Two of the three planets are in near resonance orbits with periods of 1.2d and 3.6d, while the outer planet has a period of 6.2d. All three planets are super-Earth in size with radii of \radiusb, \radiusc, and \radiusd, for \thisstar b, c, and d, respectively. At only 30 pc from the Sun, this is the closest exoplanet system discovered by the {\it K2} mission. The proximity and brightness of the host star combined with the similarity in the size of the three transiting planets make \thisstar an excellent target for comparative atmospheric characterization. The expected radial velocity semiamplitudes of the three planets are small but detectable with current instrumentation, especially given the star's fairly bright optical magnitude of V = 10.25. Radial velocity observations should be undertaken to measure the mass of each planet, to determine their interior structures for comparative studies. Mass measurements will also be critical for properly interpreting any atmospheric characterization through transit spectroscopy.

{\bf Note added in review:} During the referee process of this paper, our team became aware of another paper reporting the discovery of a planetary system orbiting \thisstar \citep{Niraula:2017}.

\acknowledgements
We thank Laura Kriedberg and Caroline Morley for their valuable conversations. Work performed by J.E.R. was supported by the Harvard Future Faculty Leaders Postdoctoral fellowship. This work was performed in part under contract with the California Institute of Technology (Caltech)/Jet Propulsion Laboratory (JPL) funded by NASA through the Sagan Fellowship Program executed by the NASA Exoplanet Science Institute. This paper includes data collected by the \Kepler/K2 mission. Funding for the \Kepler\ mission is provided by the NASA Science Mission directorate. Some of the data presented in this paper were obtained from the Mikulski Archive for Space Telescopes (MAST). STScI is operated by the Association of Universities for Research in Astronomy, Inc., under NASA contract NAS5--26555. Support for MAST for non--HST data is provided by the NASA Office of Space Science via grant NNX13AC07G and by other grants and contracts. Some of the data presented herein were obtained at the WM Keck Observatory (which is operated as a scientific partnership among Caltech, UC, and NASA). The authors wish to recognize and acknowledge the very significant cultural role and reverence that the summit of Mauna Kea has always had within the indigenous Hawaiian community. We are most fortunate to have the opportunity to conduct observations from this mountain.

\bibliographystyle{apj}

\bibliography{GJ9827}

\end{document}